\documentclass[abstracton,titlepage,12pt]{scrartcl}
\usepackage[numbers]{natbib}
\usepackage{amsmath}
\usepackage{amssymb,latexsym}
\usepackage{graphicx}
\usepackage{calc}
\usepackage[usenames]{color}
\usepackage{subfig}
\usepackage{setspace}

\usepackage[left,displaymath,mathlines]{lineno} 

\definecolor{darkred}{rgb}{1.0,1.0,0.749}
\definecolor{red}{rgb}{1.0,0.0,0.0}
\definecolor{lightred}{rgb}{0.843,0.1,0.11}
\definecolor{white}{rgb}{0.843,0.1,0.11}
\definecolor{lightblue}{rgb}{0.843,0.1,0.11}
\definecolor{blue}{rgb}{0.843,0.1,0.11}
\definecolor{darkblue}{rgb}{0.843,0.1,0.11}
\definecolor{green}{rgb}{0.13725, 0.58039, 0.11373}

\newcommand{\bc}{\begin{center}}
\newcommand{\ec}{\end{center}}
\newcommand{\bd}{\begin{description}}
\newcommand{\ed}{\end{description}}
\newcommand{\ben}{\begin{enumerate}}
\newcommand{\een}{\end{enumerate}}
\newcommand{\bi}{\begin{itemize}}
\newcommand{\ei}{\end{itemize}}
\newcommand{\be}{\begin{equation}}
\newcommand{\ee}{\end{equation}}

\renewcommand{\vec}{\mathbf}

\newcommand{\gb}{\beta}

\def\mrm#1{\mathrm{#1}}
\def\fig#1{Fig.\,\ref{#1}}

\def\sec#1{Sec.\,\ref{#1}}
\def\sect#1{Section\,\ref{#1}}
\def\eq#1{Eq.\,(\ref{#1})}

\def\ap#1{Appendix\,\ref{#1}}

\def\fig#1{Fig.\,\ref{#1}}
\begin{document}

\begin{titlepage}

\title{How to develop smarter host mixtures to control plant disease?}

\author{Alexey Mikaberidze (alexey.mikaberidze@env.ethz.ch), \\
 Bruce A. McDonald,\\
Sebastian Bonhoeffer}

\publishers{

\begin{normalsize}

\begin{flushleft}
affiliation: Institute of Integrative Biology, ETH Zurich\\
\vspace{0.5cm}
keywords: epidemiology, plant disease, mathematical model, host-pathogen
interaction, host diversity, cultivar mixture, host mixture, multiline
cultivar, population dynamics
\vspace{0.2cm}

\end{flushleft}

\end{normalsize}
}

\date{}

\end{titlepage}

\maketitle





\begin{abstract} 


  A looming challenge for agriculture is sustainable intensification
  of food production to feed the growing human population. Current
  chemical and genetic technologies used to manage plant diseases are
  highly vulnerable to pathogen evolution and are not
  sustainable. Pathogen evolution is facilitated by the genetic
  uniformity underlying modern agroecosystems, suggesting that one
  path to sustainable disease control lies through increasing genetic
  diversity at the field scale by using genetically diverse host
  mixtures. We investigate how host mixtures can improve disease
  control using a population dynamics model. We find that when a
  population of crop plants is exposed to host-specialized pathogen
  species or strains, the overall disease severity is smaller in the
  mixture of two host varieties than in each of the corresponding pure
  stands. The disease severity can be minimized over a range of mixing
  ratios. These findings may help in designing host mixtures that
  efficiently control diseases of crops. We then generalize the model
  to describe host mixtures with many components. We find that when
  pathogens exhibit host specialization, the overall disease severity
  decreases with the number of components in the mixture.
%
%
  Using these model outcomes, we propose ways to optimize the use of
  host mixtures to decrease disease in agroecosystems.

\end{abstract}

\section{Introduction}

Growing demand for higher quality food coupled with global population
growth led the Food and Agriculture Organization of the United Nations
to predict that food production will need to increase by 70\,\% by
2050 \citep{fa09}. This increase can be achieved by expanding growing
areas (extensification) or by increasing yields per hectare
(intensification). Either approach will increase the overall biomass
of crop plants and thus raise the carrying capacity of agricultural
plant pathogens. Increasing international travel and trade exposes
crops to many new pathogens, which contributes to the emergence of new
diseases. As a result of these developments, crops will become more
vulnerable to infectious diseases that cause damaging epidemics and
substantially reduce yields \citep{oe06}. 

The two most widely used disease control measures are applications of
chemicals (fungicides and antibiotics) and breeding for disease
resistant crop cultivars by incorporating resistance genes. Both of
these control measures are highly vulnerable to pathogen
adaptation. Many pathogens have repeatedly evolved to overcome
resistance conferred by major resistance genes (reviewed in
\citep{mcli02,pa02}). A recent example is the emergence of virulent
races of stem rust (called Ug99) that can infect about $90\,\%$ of
wheat varieties grown worldwide \citep{siho+11}. Similarly, many
fungicides rapidly lose their efficacy because of the emergence and
fixation of mutations encoding fungicide resistance
(e.\,g. \citep{tobr+09,brst+08}. An important disadvantage of
fungicides relative to genetic resistance is their high cost and
harmful effects on the environment. As a result of pathogen evolution,
the current commonly practiced disease control measures will likely be
inadequate to enable a sustainable intensification of food production.

Quantitative or partial resistance is thought to be more durable
\citep{pa02,pago+11}, but has not been as widely utilized as major
gene resistance. Recent research has begun to provide insights into
the molecular mechanisms responsible for quantitative resistance
\citep{poba+09,kowa10}, but studies that include quantitative
resistance in epidemiological models are rare (Lo Iacono et al.,
2012). Pathogens can still adapt to quantitative resistance leading to
an “erosion” of its effects \citep{stle+07,muco+02,mcli02,lesh97},
although at a much slower pace compared to major resistance genes.

More effective and longer-lasting disease control methods are urgently
needed to achieve a sustainable intensification of crop
production. One way to develop such methods is to focus on the
underlying properties of modern agricultural ecosystems
(agroecosystems) that make them vulnerable to plant
pathogens. Compared to natural ecosystems, agroecosystems are more
environmentally homogeneous, have a higher density of plants, and
possess much less genetic and species diversity. Increasing the
environmental and genetic homogeneity of agroecosystems enabled a high
degree of mechanization and contributed greatly to the efficiency of
food production and the development of food processing industries. But
these developments also favored the emergence of new pathogens with
higher viru¬lence and a greater degree of host specialization
\citep{stmc08} and accelerated the evolution of existing pathogens
towards higher virulence \citep{mc13,stba+11}. It is increasingly
recognized that these underlying properties of agroecosystems,
especially the lack of genetic diversity due to the dominance of
monoculture crops grown as clones, make them especially susceptible to
disease epidemics \citep{mu02,wo00,gamu99}.

For these reasons, many researchers propose to deliberately increase
genetic diversity in agroecosystems \citep{mc13,nebe+09,zhch+00} in
order to decrease disease in the short-term and enhance the durability
of disease resistance in the long-term. This diversity can be created
within a single genetic background by developing multiline cultivars
\citep{brfr69} or involve many genetic backgrounds by using variety
mixtures \citep{wo85,smle96,mu02}. Because they are based on a single,
uniform genetic background, multilines offer the advantage of a more
homogeneous crop that is more amenable to highly mechanized industrial
farming and food processing. But the disadvantage of multilines is
that it takes many years of backcrossing to create a multiline
variety, and the resulting multiline is unlikely to possess the best
yield or quality characteristics for all local environments. But
progress in genetic engineering of plants is likely to lead to the
development of resistance gene cassettes \citep{wuho+11,mc13} that can
enable the rapid synthesis of locally adapted multiline cultivars that
carry an assortment of different resistance genes, combining the
resistance gene diversity useful for controlling diseases with the
background crop genotype uniformity useful for efficient food
production and processing. In this study the distinction between
multiline cultivars and variety mixtures is not important, so we will
refer to both options simply as host mixtures.

Many field experiments have been performed to determine whether host
mixtures reduce the amount of fungal disease on crop plants
(e.g. \citep{husu+12,nish+12,negu11,comu02,zhch+00,neel+97,muha+94,chwo84},
see also reviews \citep{waav+12,mu02,figa+00,smle96,wo85} and
references therein). The findings of over 30 studies (mostly in
barley, wheat, rice and beans) were summarized in \citep{smle96}. The
vast majority of experiments showed less disease in mixtures as
compared to the mean of the pure stands for obligate pathogens such as
rusts and mildews. However, there was a large variation in the
percentage of disease reduction: for example, between 9\,\% and 80\,\%
for powdery mildew in barley, and between 13\,\% and 97\,\% for stripe
rust in wheat. A recent meta-analysis of stripe rust on wheat
considered 161 mixture cases reported in 11 publications
\citep{husu+12}. In 83\,\% of these cases the average disease level
was found to be lower in mixtures compared to the mean of the pure
stands. A reduction in disease of between 30\,\% and 50\,\% was found
most frequently. A large-scale study performed in China demonstrated
that row mixtures of rice varieties could strongly reduce rice blast
\citep{zhch+00}. Thus, host mixtures reduce the amount of disease
in most studied cases, but the outcomes exhibit a wide variation, even
within a single study (for example \citep {comu02}).

This variation is one of the reasons why multilines and cultivar
mixtures have so far gained little acceptance among seed companies or
growers. To achieve reliable disease control, we need to identify the
conditions under which mixtures work best and use this knowledge to
design optimal mixtures. This requires a better understanding of the
underlying mechanisms of disease reduction in mixtures. Our study
contributes to this understanding in three important ways by using a
population dynamics model of plant-pathogen interactions. First, we
identified conditions where mixtures are superior compared to pure
stands. Second, we defined optimal ratios of components to include in
the mixture. Third, we determined optimal numbers of components to
include in the mixture.

This was done by exploring possible disease outcomes when two or more
hosts are mixed in the presence of two or more pathogen strains or
species. This approach is a substantial advance with respect to
previous modeling studies that mostly concentrated on a single
pathogen strain \citep{saty+10,skro+05,gamu99,bove+90,mubr88} with the
exception of a few studies that considered several pathogen genotypes
(e.\,g. \citep{xuri00,li10,lava+95}). It is crucial to consider more
than one pathogen in the model, because practically all fields are
colonized by more than one pathogen strain and several pathogen
species and the full advantage of host mixtures manifests in reduction
of the overall pathogen load. Moreover, we obtained analytical
solutions that allowed us to investigate the disease reduction over
the whole range of parameters that includes both qualitative and
quantitative host resistance (see \sec{sec:transm-matr-val}).

\section{Model and assumptions}
\label{sec:model}

\begin{figure}
  \centerline{\includegraphics[width=0.9\textwidth]{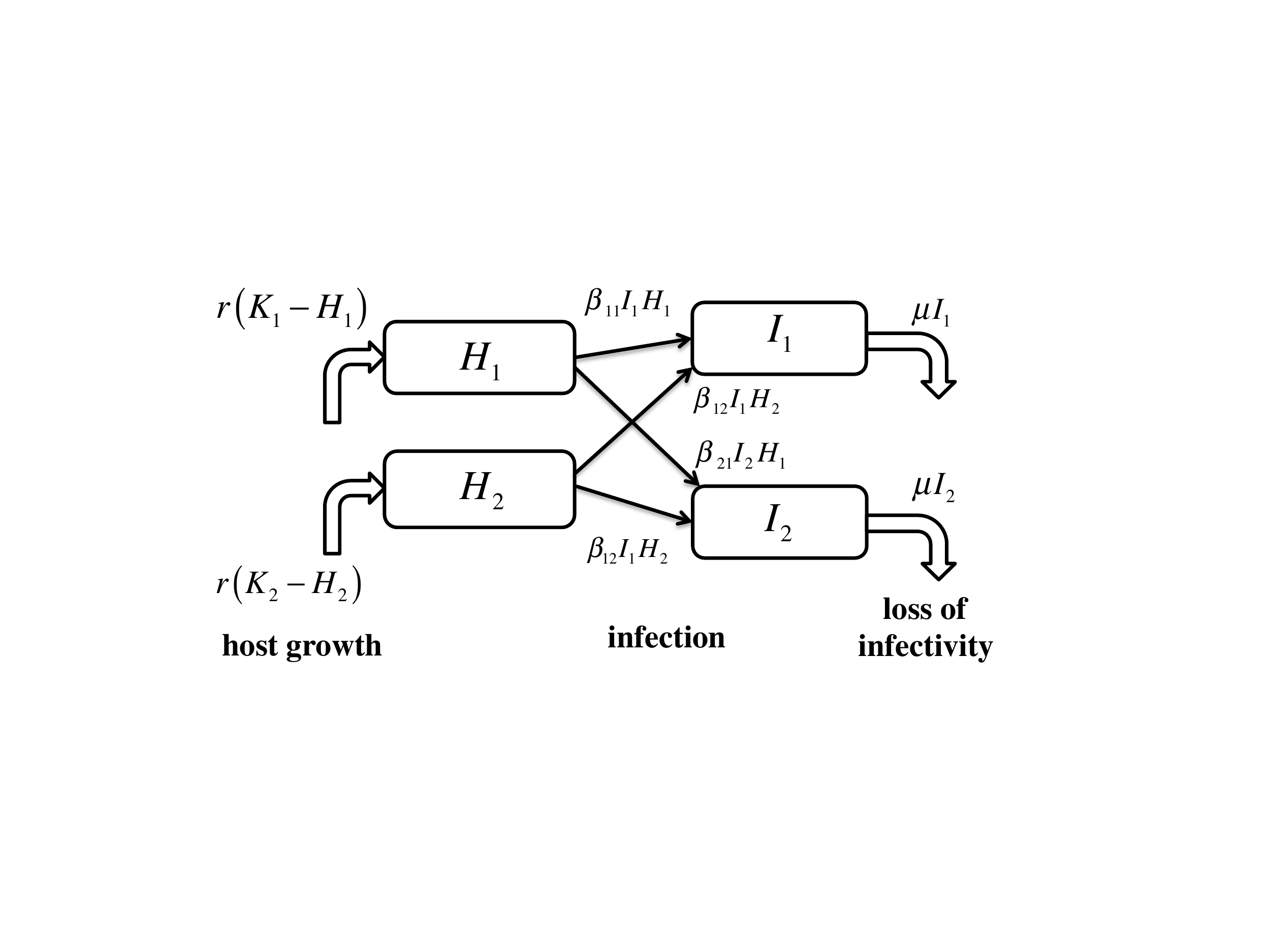}} \caption{
Sheme of the model equations (\ref{eq:2host2fung-1})-(\ref{eq:2host2fung-4}).}
\label{fig:model-scheme}
\end{figure}

We first consider the case when two host varieties $H_1$ and $H_2$ are
exposed to two types of pathogen: 1 and 2 (we also refer to them as $P_1$
and $P_2$). These could be either different strains (races or pathotypes) of the same
pathogen or different pathogen species capable of infecting the same
host tissue.
To describe the host-pathogen interaction, we use a deterministic
epidemiological model of susceptible-infected dynamics (see
\fig{fig:model-scheme}), which is an extension of the model described
previously \citep{mimc+13} to the case of two different hosts. This
model can be applied to a variety of aerially and splash-dispersed,
polycyclic pathogens of cereal crops, such as the fungi and bacteria
causing rusts, mildews, blasts, spots and blotches.
\begin{linenomath}
\begin{align}
\frac{d H_1}{d t} &= r ( K_1 - H_1) - (\beta_{11} I_1 + \beta_{21} I_2) H_1, \label{eq:2host2fung-1}\\ 
\frac{d H_2}{d t} &= r (K_2 - H_2) - (\beta_{12} I_1 + \beta_{22} I_2) H_2, \label{eq:2host2fung-2}\\ 
\frac{d I_1}{d t} & = (\beta_{11} H_1 + \beta_{12} H_2) I_1 - \mu I_1, \label{eq:2host2fung-3}\\ 
\frac{d I_2}{d t} & = (\beta_{21} H_1 + \beta_{22} H_2) I_2 - \mu  I_2. \label{eq:2host2fung-4}
\end{align}
\end{linenomath}
There are four compartments in the model: susceptible hosts $H_1$ of
variety 1, susceptible hosts $H_2$ of variety 2, hosts $I_1$ infected
by pathogen 1 and hosts $I_2$ infected by pathogen 2. Each of the
quantities $H_1$, $H_2$, $I_1$, $I_2$ represents the total amount of
the corresponding host tissue within one field, which could be leaves,
stems or grain tissue, depending on the host-pathogen
combination. 

Susceptible hosts $H_1$ and $H_2$ grow with the same rate $r$. Their
growth is limited by their ``carrying capacities'' $K_1$ and $K_2$,
implying limitations in space or nutrients. 
One can vary the proportion of host plants of the two varieties by
adjusting the ratio of the corresponding seeds to be planted. This is reflected
in the change of the ratio $\phi_1 = K_1/(K_1+K_2)$ in the
model. We assume that the seeds of the two host varieties are well
mixed before planting, such that the spatial distribution across the
field is uniformly random for both types of plants.
The infected host tissue loses its infectivity (i.\,e. the ability to
produce infectious spores) with the rate $\mu$ ($\mu^{-1}$ is the
average infectious period), which is assumed to be
the same for $I_1$ and $I_2$.
%

\begin{figure}
  \centerline{\includegraphics[width=0.4\textwidth]{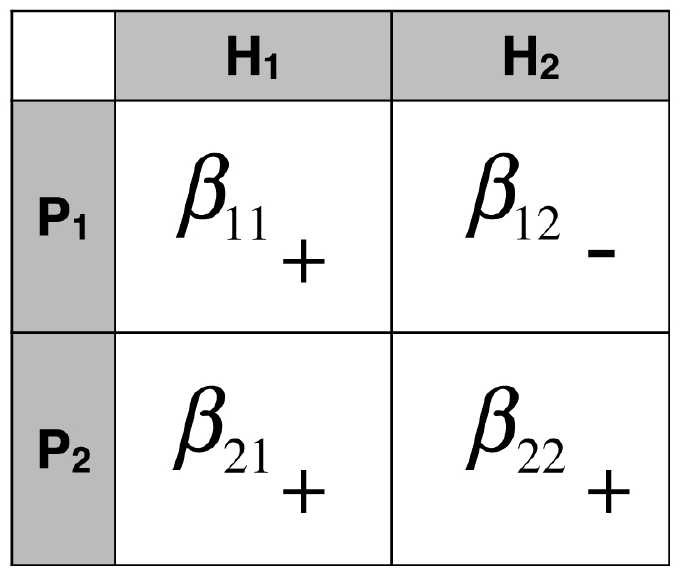}} \caption{
    Scheme of the host-pathogen interaction. ``+", ``--" signs
    correspond to a ``pure'' gene-for-gene (GFG) interaction. The
    transmission matrix $\beta_{ij}$, $i,j=1,2$ represents a more general
    description with ``pure'' GFG ($\beta_{11}=\beta_{22}=\beta_{21}>0$; $\beta_{12}=0$) and
    full host specialization ($\beta_{11},\beta_{22}>0$; $\beta_{12}=\beta_{21}=0$) as
    limiting cases.}
\label{fig:gfg-diagr}
\end{figure}

We assume that the two host varieties differ only in their
susceptibility to the two pathogens, and the two pathogens differ only
in their capability to infect the two hosts, which is reflected in the
rate of spore production and the ability of resulting spores to infect
additional host tissue.
%
%
Both host susceptibility and pathogen virulence are described in the
model by the four transmission rates $\beta_{11}$, $\beta_{22}$,
$\beta_{12}$, and $\beta_{21}$.  They can be conveniently arranged in a
transmission matrix or WAIFW (Who Acquires Infection From Whom) matrix
\begin{equation}\label{eq:transm-matr}
\vec{B} = \left| \begin{array}{cc} \beta_{11} & \beta_{12} \\ \beta_{21} & \beta_{22} \end{array} \right|.
\end{equation}
The first index of matrix elements represents the source of infection
and the second index represents the recipient of infection (see \fig{fig:gfg-diagr}). For
example, $\beta_{12}$ describes the transmission rate from $I_1$ to
$H_2$. Possible relationships between the elements of the transmission
matrix are discussed in \sect{sec:transm-matr-val}.
We neglected spatial dependence of pathogen dispersal: every infected
host is equally likely to infect every other infected host within the
population (often called the ``mass-action'' approximation). This approximation is
valid for air-borne pathogens with long-range dispersal (for example,
rusts and mildews) and for sufficiently small plot sizes.

Note that $I_1$ includes the tissue of both hosts infected by pathogen 1. Similarly, $I_2$ includes host tissue of both hosts
infected by the pathogen 2. This formulation assumes that the
transmission rate does not depend on the host variety of the source of
infection, but only depends on the host variety of the recipient of
infection. 
In other words, under this assumption, the spore production rate and
the quality of spores produced depend on the pathogen genotype, but
not on the host genotype. But the infection efficiency (or infection
success) of a spore depends on the host genotype on which it lands.
%
In order to relax this assumption, one needs to subdivide
each of $I_1$ and $I_2$ into two compartments, according to the type
of host tissue infected.
%

In order to quantify the amount of disease, we use the proportion of
infected area of the host tissue
\begin{equation}\label{eq:beta-def}
y = (I_1 + I_2)/(I_1 + I_2 + H_1 + H_2),
\end{equation}
and call it \emph{disease severity}. We will often use the
equilibrium values of $y$, $I_1$, $I_2$, $H_1$, $H_2$ (denoted by
an ``*''-superscript), which are achieved over long periods of time. They
correspond to the stable fixed point of the system
(\ref{eq:2host2fung-1})-(\ref{eq:2host2fung-4}), as explained in the
\ap{apsec:fps} and \ref{apsec:2h2p-linstab}.

We will also consider a more general case of a mixture with $n$ hosts
that is exposed to $n$ pathogens. In this case, the dynamics of
the host-pathogen interactions are described by this system of $2 n$
equations that is a generalization of the system of
Eqs.\,(\ref{eq:2host2fung-1})-(\ref{eq:2host2fung-4}):
\begin{linenomath}
\begin{align}
\frac{d H_i}{d t} &= r ( K_i - H_i) - \sum_{k=1}^{n} \beta_{ki} I_k H_i, \label{eq:hi-nhost-npath}\\ 
\frac{d I_i}{d t} & = \sum_{k=1}^{n} \beta_{ik} I_i H_k - \mu  I_i,\: i=1,...,n \label{eq:ii-nhost-npath}
\end{align}
\end{linenomath}
Here, the transmission matrix $\vec{B}$ is an $n \times n$ square matrix, the
element $\beta_{ik}$ describes the transmission rate of the pathogen that
originates from the infected host of type $i$ and infects the healthy
host of variety $k$. We assume that every host variety is planted at
the same proportion, i.\,e. $K_i = K$. 

We will vary the number of host varieties in the mixture $n$,
while keeping the total carrying capacity constant: $K_\mrm{tot} =
\sum_{i=1}^n K_i = n K$. We will consider the total amount of healthy
and infected hosts at the infected equilibrium of the system of
Eqs.\,(\ref{eq:hi-nhost-npath})-(\ref{eq:ii-nhost-npath})
\begin{equation}\label{eq:htot-itot-def}
H_\mrm{tot}^* = \sum_{i=1}^n H_i^*, \: I_\mrm{tot}^* = \sum_{i=1}^n I_i^*.
\end{equation}
and the total disease severity defined by
\begin{equation}\label{eq:sevtot-def}
y_\mrm{tot}^* = \frac{I_\mrm{tot}^*} {I_\mrm{tot}^* + H_\mrm{tot}^*}.
\end{equation}

In order to obtain an analytical solution for the disease severity
\eq{eq:sevtot-def}, we consider the transmission matrix of a simple
form
\begin{linenomath}
\begin{equation}\label{eq:bmatr-ps}
\vec{B} = \begin{pmatrix}
\beta_\mrm{d} & \beta_\mrm{nd} & \cdots & \beta_\mrm{nd}\\
\beta_\mrm{nd} & \beta_\mrm{d} &  \cdots & \beta_\mrm{nd}\\
\vdots &  & \ddots &\\
\beta_\mrm{nd} & \cdots & & \beta_\mrm{d}
\end{pmatrix}
\end{equation}
\end{linenomath}
Here, every diagonal element of the matrix $\vec{B}$ is equal to
$\beta_\mrm{d}$ and every non-diagonal element is $\beta_\mrm{nd}$. We
generally assume partial specialization, where $\beta_\mrm{d} \geq
\beta_\mrm{nd}$. Furthermore, assuming that all healthy and infected hosts
start with the same initial conditions, their dynamics will be the
same.  Hence, $H_i=H_p$, $I_i=I_p$ for any $i$ and we can simplify the
Eqs.\,(\ref{eq:hi-nhost-npath})-(\ref{eq:ii-nhost-npath}):
\begin{linenomath}
\begin{align}
\frac{d H_p}{d t} &= r ( K - H_p) - \beta_\mrm{eff} I_p H_p, \label{eq:hi-nhost-npath-ps}\\ 
\frac{d I_p}{d t} & =  \beta_\mrm{eff} I_p H_p - \mu  I_p, \label{eq:ii-nhost-npath-ps}
\end{align}
\end{linenomath}
where $\beta_\mrm{eff} = \beta_\mrm{d} + (n-1) \beta_\mrm{nd}$.

\section{Possible relationships between transmission rates}
\label{sec:transm-matr-val}

What are the possible relationships between the transmission rates
$\beta_{ij}$, $i,j=1,2$? The answer requires an understanding of the
host-pathogen interaction on the molecular level
\citep{bema07,joda06}.
Our current knowledge can be summarized in a simplified way using a four-stage model
\citep{daho+13,bema07,joda06}. First, plants have a basal (or innate) immune
system. It consists of pathogen recognition proteins (PRPs) that 
respond to microbe-associated molecular patterns (MAMPs, also called
pathogen-associated molecular patterns or PAMPs). MAMPs are typically
highly conserved molecules produced by pathogens, for example
chitin in fungi or flagellin in bacteria.
Upon MAMP recognition, PRPs activate basal immune responses and the
infection is prevented. Second, pathogens evolve to suppress
the basal immune responses by producing multiple effector proteins (E-proteins or effectors), which target host proteins and suppress host
resistance. Third, plant resistance genes (R-genes) produce NB-LRR proteins
(R-proteins), which recognize the effector proteins of the pathogen
and restore resistance. Finally, pathogens avoid recognition by
modifying or removing the effector proteins. As a result, the R-proteins can no longer recognize the effector proteins and resistance
is lost again. Thus loss or modification of effector proteins is
expected to confer a fitness cost to pathogens, since they can be
essential for pathogenicity in the second step described
above. 
%
Loss of some effectors (called ``core effectors'') confers a
sizable fitness cost. These effectors can be identified using a
combination of genetic and genomic methods \citep{daho+13}.

Next, we discuss how different possibilities of host-pathogen
interaction on the molecular level affect the relationship between the
elements of the transmission matrix $B$.  First, assume that only one
combination of an R-protein and an E-protein determines the values of
the transmission matrix. Consider the case when host 1 has the
R-protein, while host 2 does not have it; and pathogen 1 has the corresponding E-protein, while pathogen 2 does not have it.
In this case, the interaction follows the ``pure'' GFG scheme, i.\,e.
$\beta_{11}=\beta_{22}=\beta_{21}>0$, $\beta_{12}=0$ (see \fig{fig:gfg-diagr}),
provided that elimination the E-protein does not affect pathogen
fitness and the presence of the R-gene does not confer a fitness cost
to the host. We denote this scenario as (A).
If the loss or mutation of an effector confers a fitness cost, then
$\beta_{11}>\beta_{21}$. Thus, we obtain $\beta_{11}>\beta_{22}=\beta_{21}>0$, $\beta_{12}=0$
and denote this scenario as (B).
In the case when the presence of an R-protein confers a fitness
cost to the host that manifests in an increased susceptibility to
the pathogen $P_2$, then we have the relationship
$\beta_{11}>\beta_{22}>\beta_{21}>0$, $\beta_{12}=0$, and call it scenario (C).


The three scenarios considered above result from an interaction of a
single R-protein -- E-protein pair (R--E pair), but in reality there
are many R-proteins and many E-proteins \citep{bema07}, which can be
active in the same host-pathogen combination.
Effects of these multiple interactions on the host susceptibility to
disease are expected to sum up \citep{bema07}.
%
%
Under the GFG scheme [scenario (A) above], their combined effect may
manifest as a
``pure'' GFG. Alternatively, it can manifest as an interaction
with a quantitative degree of specialization. 
%
%
%
This can occur when some of the R-proteins are present in host 1, but absent
in host 2 and vice versa, and, similarly, some of E-proteins are
present in pathogen 1, but absent in pathogen 2 and vice versa.

To better understand this, consider the simplest example when there
are only two R--E pairs, R1--E1 and R2--E2, where R1 only recognizes
E1 and R2 only recognizes E2. Complete resistance occurs when both
recognition events occur, i.\,e. when the host
has both R1 and R2 and the pathogen has both E1 and E2.
Consider the case when each pathogen has one of the effectors, but
lacks the other one; and each host has one of the R proteins, but
lacks the other one. For example, pathogen 1 has E1, but lacks E2,
while pathogen 2 has E2, but lacks E1. Similarly, host 1 has R2, but
lacks R1, while host 2 has R1, but lacks the R2. In this case, the
infection of host 2 by pathogen 1 (transmission rate $\beta_{12}$) is
suppressed by the recognition of the protein E1 by the protein
R1. Here, only one of the two possible R--E recognitions occur,
consequently the resistance is incomplete.
%
%
The same happens when host 1 is attacked by pathogen 2 (rate $\beta_{21}$). In
the other two possibilities, when pathogen 1 attacks host 1, or
pathogen 2 attacks host 2, we expect full susceptibility, since R--E
recognition is absent. Thus, pathogen 1 specializes on host 1, and
pathogen 2 specializes on host 2, but the degree of specialization is
incomplete, since the cross-transmission rates are still
positive. This scenario [we will call it (D)] corresponds to both
diagonal elements of the transmission matrix being larger than both
non-diagonal ones, i\,e. $\beta_{11}, \beta_{22}>\beta_{21},\beta_{12}>0$.

Moreover, if we allow for further deviations from the ``pure'' GFG scheme due
to fitness costs of losing effectors for pathogens and also costs of
having ``unnecessary'' R-proteins for hosts, then many more outcomes
become possible and we are even more likely to find pairs of pathogens and
hosts which exhibit a degree of specialization
[i.\,e. interact via scenario (D) discussed above].

As we have seen above, the variety of possible outcomes for the
transmission matrix $\vec{B}$ is already quite large even when
considering only two R-proteins and two E-proteins. For larger
numbers of R- and E-proteins, we expect the transmission matrix
$\vec{B}$ to exhibit even richer behavior.
%
%
Therefore, it is desirable to study the benefit of mixing host
varieties representing the whole range of values of the matrix
elements of $\vec{B}$. We have done this by obtaining analytical
expressions for the disease severity and frequencies of pathogens as
functions of the matrix elements $\beta_{ij}$ and other model
parameters (see \ap{apsec:fps}). This is an advantage of our study
with compared to previous theoretical investigations that assumed a
``pure GFG'' interaction, without fitness costs associated with losing
effectors \citep{ohsa06,bogi03,iabo+13}, or that assumed full
specialization \citep{li10}, where each pathogen can only infect its
preferred host and is unable to infect any other hosts (also called
the ``matching alleles'' model \citep{kili12}). The latter scenario seems to
represent only a hypothetical limiting case, because it requires full
resistance, which is unlikely given the simultaneous
presence of many pairs of R- and E-proteins.
%
%
In contrast, partial specialization (scenario (D)), when the diagonal
elements $\beta_{11}$ and $\beta_{22}$ are larger than non-diagonal
ones $\beta_{12}$ and $\beta_{21}$, but the non-diagonal ones are
still significantly larger than zero, seems to be the most generic
case. This is because it arises from a ubiquitous GFG-type of
interaction with many R-proteins present in the host, many
corresponding E-proteins present in the pathogen, as well as fitness
costs for the pathogen due to elimination or modification of
E-proteins and fitness costs for the host due to having unnecessary
R-proteins.




Our conclusion that partial specialization is quite common, if not
universal in plant-pathogen interaction is confirmed by the findings
of several artificial selection experiments
\citep{le69,chwo84,vila00,zhmu+02,zhmc13}, where pathogen strains were observed that
are better adapted to particular host cultivars than to other
cultivars (differential adaptation). In several cases, quantitative
resistances were identified in these host cultivars
\citep{lesh97,kole86}. Also, cross-inoculation studies with field
isolates demonstrate that pathogen strains are often better adapted to
cultivars from which they were isolated \citep{ahmu+95,anpi+07}. [See
\citep{para+09} p.\,417-419 for a detailed discussion]. Moreover,
local adaptation of pathogens to their hosts was detected in studies
of wild plant-pathogen systems (e.\,g. \emph{Linum
  marginale--Melampsora lini} \citep{thbu+02}, \emph{Plantago
  lanceolata--Podosphaera plantaginis} \citep{la07}).






\section{Results}
\label{sec:results}
%


We first consider the effect of host mixtures on the competition
between the two pathogen strains (\sec{sec:path-comp}). Then, in
\sec{sec:optmix} we determine what proportions of hosts in the mixture
will minimize the disease. Finally, we generalize our approach to the
case of many pathogen strains and host varieties and determine an
optimal number of components in a host mixture (\sec{sec:optncomp}).


\subsection{Long-term outcomes of the host-pathogen dynamics}
\label{sec:path-comp}


Understanding long-term outcomes is crucial to determine whether a
host mixture will reduce the amount of disease.
%
%
Moreover, long-term outcomes are simple to determine, because they
correspond to stability ranges of the fixed points of the system of
Eqs.\,(\ref{eq:2host2fung-1})-(\ref{eq:2host2fung-4}) and can be found
analytically (see Appendix\,\ref{apsec:fps} and \ref{apsec:2h2p-linstab}).

\begin{figure}
  \centerline{\includegraphics[width=0.9\textwidth]{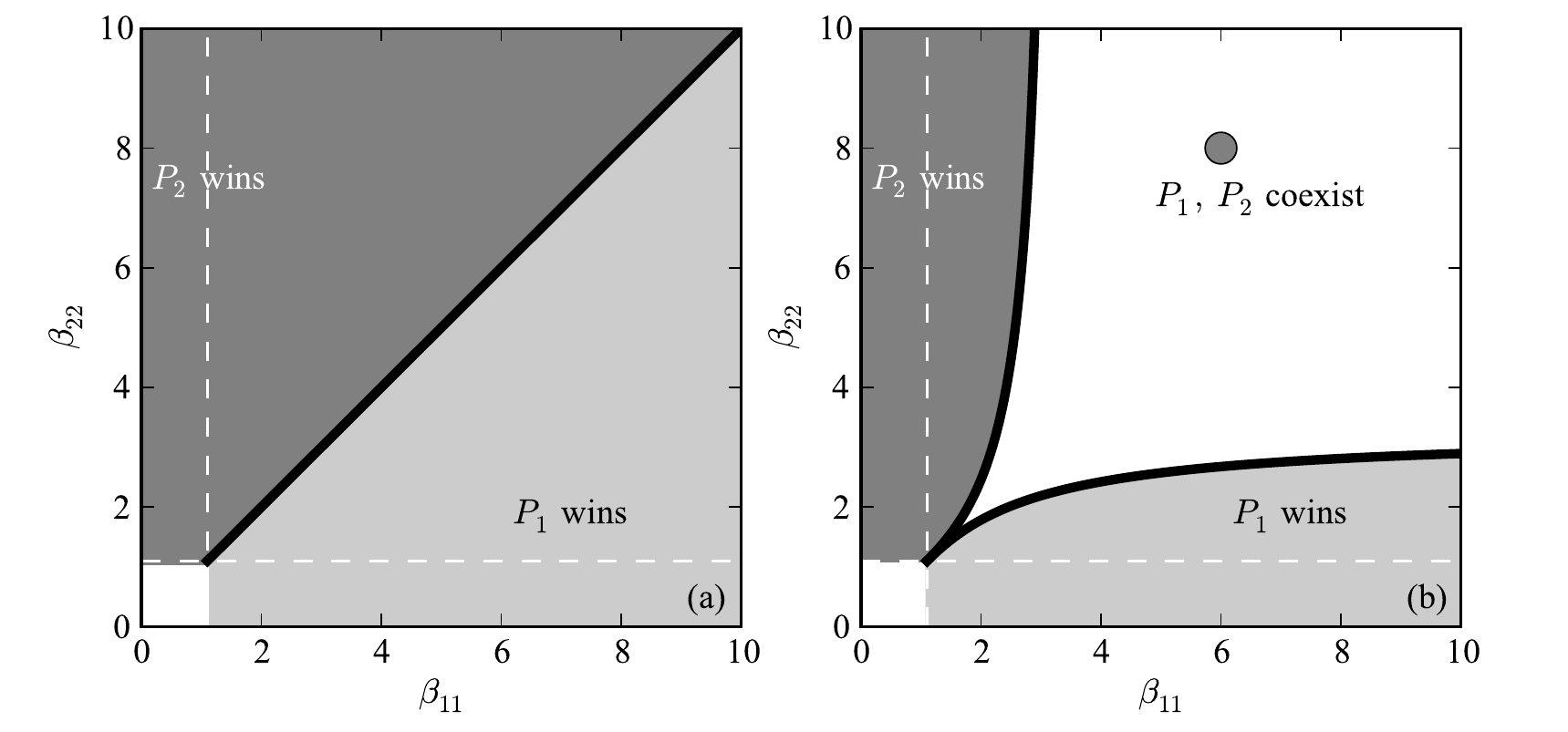}} \caption{
    Long-term outcomes of the host-pathogen dynamics described by
    Eqs.\,(\ref{eq:2host2fung-1})-(\ref{eq:2host2fung-4}) versus the
    transmission rates $\beta_{11}$ and $\beta_{22}$. Panel (a): both
    host varieties have the same susceptibility,
    i.\,e. $\beta_{11}=\beta_{12}=\gb_1$,
    $\beta_{22}=\beta_{21}=\gb_2$; panel (b): the two host varieties
    differ in their susceptibility ($\beta_{11}$ and $\beta_{22}$ are
    varied, while $\beta_{12} = 0.9$, $\beta_{21}=0.9$). Possible
    outcomes correspond to different fixed points of
    Eqs.\,(\ref{eq:2host2fung-1})-(\ref{eq:2host2fung-4}) and their
    ranges coincide with the ranges of stability of the fixed points
    (see Appendix\,\ref{apsec:fps} and \ref{apsec:2h2p-linstab}). They
    are shown in different shades of grey: (i) both pathogens $P_1$ and $P_2$
    die out (white rectangle in the lower left corner); (ii) $P_1$
    survives, $P_2$ dies out (dark grey); (iii)
    $P_2$ survives, $P_1$ dies out (light grey); (iv) $P_1$ and $P_2$
    coexist (white). Solid black curves in (b) are plotted according
    to \eq{eqap:b11c} and \eq{eqap:b22c}. Parameter values: $K_1=K_2 = 0.5$,
    $\mu=1$. Grey circle corresponds to same parameter values as
    dashed vertical lines in \fig{fig:pd-freqres-vs-a1-somespec}.}
\label{fig:pd-h1h2-small-cross-infect}
\end{figure}

The case when the two host varieties have the same susceptibility to
disease, $\beta_{11}=\beta_{12}=\gb_1$, $\beta_{22}=\beta_{21}=\gb_2$ is equivalent to
having just one host variety. In this case, the long-term outcome
depends on the basic reproductive numbers of the two pathogens:
$R_{01} = \gb_1 K_1 / \mu$ and $R_{02} = \gb_2 K_2 / \mu$. If both
$R_{01}<1$ and $R_{02}<1$, then both pathogens die out [gray region in
\fig{fig:pd-h1h2-small-cross-infect}(a)]. If at least one of the $R_0$'s
exceeds unity, then the pathogen with the larger $R_0$ survives and
the pathogen with the smaller $R_0$ dies out, and no stable
co-existence is possible, as shown by red and blue regions in
\fig{fig:pd-h1h2-small-cross-infect}(a).


In contrast, when the two host varieties differ in their
susceptibility to disease, stable co-existence of the two pathogens
becomes possible. This is realized when each of the pathogens at
least partially specializes on one of the hosts, i.\,e. $\beta_{11},
\beta_{22}>\beta_{21},\beta_{12}$ [scenario (D) discussed in \sec{sec:model}]. Now,
the diagram of outcomes looks different
[cf. \fig{fig:pd-h1h2-small-cross-infect}(a) and (b)]: the regions of
domination of $P_1$ and $P_2$ are now separated by a broad region,
where $P_1$ and $P_2$ exhibit stable co-existence.



The vertical dashed line indicates the threshold value
$\beta_{22}=\beta_\mrm{22c}$ [\eq{eqap:b22c}], above which pathogen 1 can
invade the host population in the absence of pathogen
2. Similarly, the horizontal dashed line shows the invasion threshold
$\beta_{11}=\beta_\mrm{11c}$ [\eq{eqap:b11c}] of pathogen 2 in the absence
of pathogen 1. Hence, in the lower left region of the graph,
separated by these two lines, neither pathogen can invade and
therefore they die out.

At $\beta_{11}>\beta_\mrm{11c}$ pathogen 1 can invade the host
population in the absence of pathogen 2. Assume that this has happened
and pathogen 1 has reached equilibrium with the host population. Note
that the threshold value of $\beta_{22}$, above which pathogen 2 can
also invade is larger than in the absence of pathogen 1. This is
because pathogen 1 has occupied some of the healthy host tissue, and
since the basic reproductive number is proportional to the amount of
healthy host tissue, the threshold value $\beta_{22c}$ increases.
%
%
Thus, if there is a degree of specialization of the pathogens with
respect to their hosts (i.\,e. $\beta_{11},
\beta_{22}>\beta_{12},\beta_{21}$), then the stable coexistence of the
two pathogens is possible. 



\subsection{Is there an optimal mixture of host varieties?}
\label{sec:optmix}


Planting a mixture of host varieties provides an additional
parameter that can be adjusted, namely the proportions of the varieties in the
mixture. Does planting a mixture of hosts reduce the total amount of
disease compared to the case of monoculture stands? Furthermore, is
there an optimal proportion of the host varieties at which the
amount of disease is minimized?
Answers to these questions depend on the relationships between the
elements of the transmission matrix $\vec{B}$.

\begin{figure}
  \centerline{\includegraphics[width=0.5\textwidth]{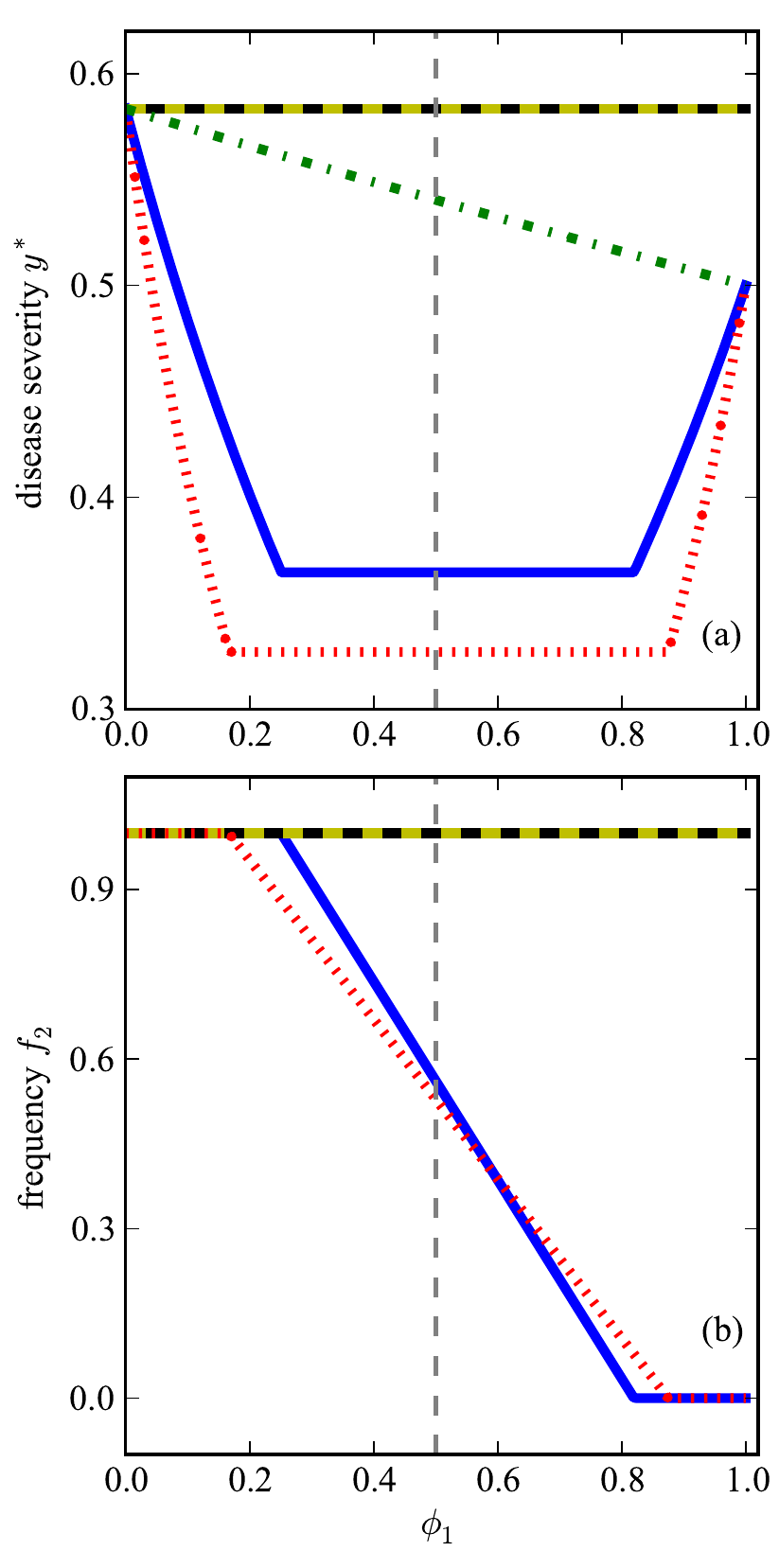}} \caption{
    Disease severity $y^*$ (upper panel) and the frequency $f_2^*=I_2^*/(I_1^*+I_2^*)$ of
    pathogen 2 (lower panel) at equilibrium as functions of
    the proportion of host 1 in the mixture
    $\phi_1=K_1/(K_1+K_2)$, according to Eqs.\,(\ref{eq:pdis-eq-app}),
    (\ref{eq:f2-def}). Parameter values: $\beta_{11} = 6$, $\beta_{22}= 8$,
    $K=K_1+K_2 = 1$, $r=0.2$, $\mu=1$. The non-diagonal elements of
    the infection matrix $\vec{B}$ determine the degree of
    specialization: (a) full specialization $\beta_{12} = \beta_{21}=0$ (red
    dotted); (b) small degree of specialization $\beta_{12} = \beta_{21}=0.9$ (blue,
    solid); (c) no specialization $\beta_{12} = \beta_{11}=6$,
    $\beta_{21}=\beta_{22}=8$ (black, upper); (d) strict gene-for-gene
    interaction $\beta_{21} = \beta_{22}=8$, $\beta_{12}=0$ (yellow, upper); (e)
    single pathogen $\beta_{11} =\beta_{21} = 6$, $\beta_{22} = \beta_{12}=8$
    (green, dash-dotted). Cases (c) and (d) correspond to the upper lines
    and overlap completely.}
\label{fig:pd-freqres-vs-a1-somespec}
\end{figure}

We calculate the disease severity at equilibrium $y^{*}$
[\eq{eq:pdis-eq-app}] as a function of the proportion of the host variety
1 in the mixture $\phi_1=K_1/K$ [see
\fig{fig:pd-freqres-vs-a1-somespec}(a)]. The quantity $\phi_1$ is varied
from zero to one, while keeping the total carrying capacity of hosts
$K = K_1 + K_2$ constant.

When each pathogen can infect both hosts equally well
(i.\,e. $\beta_{12}=\beta_{11}$, $\beta_{21}=\beta_{22}$, no specialization), disease
severity does not depend on $\phi_1$ [black dashed curve in
\fig{fig:pd-freqres-vs-a1-somespec}(a)]. The same outcome is observed
when the host-pathogen interaction strictly follows the gene-for-gene
scheme, i.\,e. $\beta_{21} = \beta_{22}>\beta_{11}>0$, $\beta_{12}=0$ [yellow dashed
curve in \fig{fig:pd-freqres-vs-a1-somespec}(a)]. We used the values
of the transmission rates, which satisfy $\beta_{22}>\beta_{11}$. Hence, 
pathogen 2 is fitter than pathogen 1 and dominates the population
and at any value of $\phi_1$ [black and yellow dashed curve in
\fig{fig:pd-freqres-vs-a1-somespec}(b)].

In the case of a single pathogen infecting a mixture of hosts with
different degrees of susceptibility ($\beta_{22}=\beta_{12}>\beta_{11}=\beta_{21}$),
the disease severity decreases linearly with $\phi_1$. In this case, simply
using a monoculture with the more disease-resistant host variety
($\phi_1=1$) would reduce the disease most strongly [green dashed-dotted
curve in \fig{fig:pd-freqres-vs-a1-somespec}(a)]. This is in agreement
with findings of an experiment, in which a mixture of a susceptible
and resistant barley variety was infected by barley powdery mildew
(caused by \emph{Blumeria graminis} f. sp. \emph{hordei}) reported in
\citep{figa+00}. In this study the disease reduction was found to
decrease linearly with the proportion of the susceptible variety in
the mixture.

The picture changes if there is a degree of specialization of pathogen
strains or species to host varieties ($\beta_{22},\, \beta_{11} >\beta_{12},\, \beta_{21}$). In this case the disease severity $y^*$ first
decreases with $\phi_1$, then reaches a constant value, and after that
increases again. Thus, the disease is reduced over a range of
intermediate values of $\phi_1$ (solid red and blue curves).
The magnitude of this reduction increases with the degree of
specialization and reaches a maximal value at full specialization
(solid red curve). Also, the range of $\phi_1$-values, over which the
proportion of disease remains minimal, increases with the degree of
specialization [cf. blue and red curves in
\fig{fig:pd-freqres-vs-a1-somespec}(a)].

The ranges over which the frequency of pathogen 2 remains constant or
changes as a function of the cropping ratio $\phi_1$ correspond to the
ranges of stability of different fixed points of the model system
Eqs.\,(\ref{eq:2host2fung-1})-(\ref{eq:2host2fung-4}). This can be
seen from \fig{fig:pd-freqres-vs-a1-somespec}(b), where the frequency
$f_2$ of pathogen 2 is shown versus $\phi_1$. In the region where
$y^*$ decreases with $\phi_1$, pathogen 2 dominates the population
($f_2=1$). In the region where $y^*$ stays constant, the two pathogens
co-exist, but the frequency of pathogen 2 decreases with $\phi_1$
until it reaches zero. This occurs at the border, where another fixed
point becomes stable, the one corresponding to pathogen 1 dominating
the population ($f_2=1$). Here, the disease severity increases with $\phi_1$.

Why does the disease severity decrease with $\phi_1$ at small values of
$\phi_1$? In this parameter range, pathogen 2 dominates the
population in the long term. Since pathogen 2 specializes on
host 2, it develops best when only host 2 is planted, i.\,e. at
$\phi_1=0$. By adding a small amount of host 1 to the mixture, we create
suboptimal conditions for pathogen 2: it is still able to
outcompete pathogen 1, but since there is less of its preferred
host tissue, the resulting disease severity is smaller. A similar
explanation holds for the increase of disease severity with $\phi_1$ at
large values of $\phi_1$.


Why does the disease severity stay constant over a range of
intermediate values of $\phi_1$? This range corresponds to co-existence
of the two pathogens. Since there is a degree of specialization, by
increasing $\phi_1$ we make pathogen 1 more fit while pathogen 2
becomes less fit.  These two changes compensate each other, so that
the total disease severity, which includes both pathogen strains,
remains the same.

\begin{figure}
  \centerline{\includegraphics[width=0.6\textwidth]{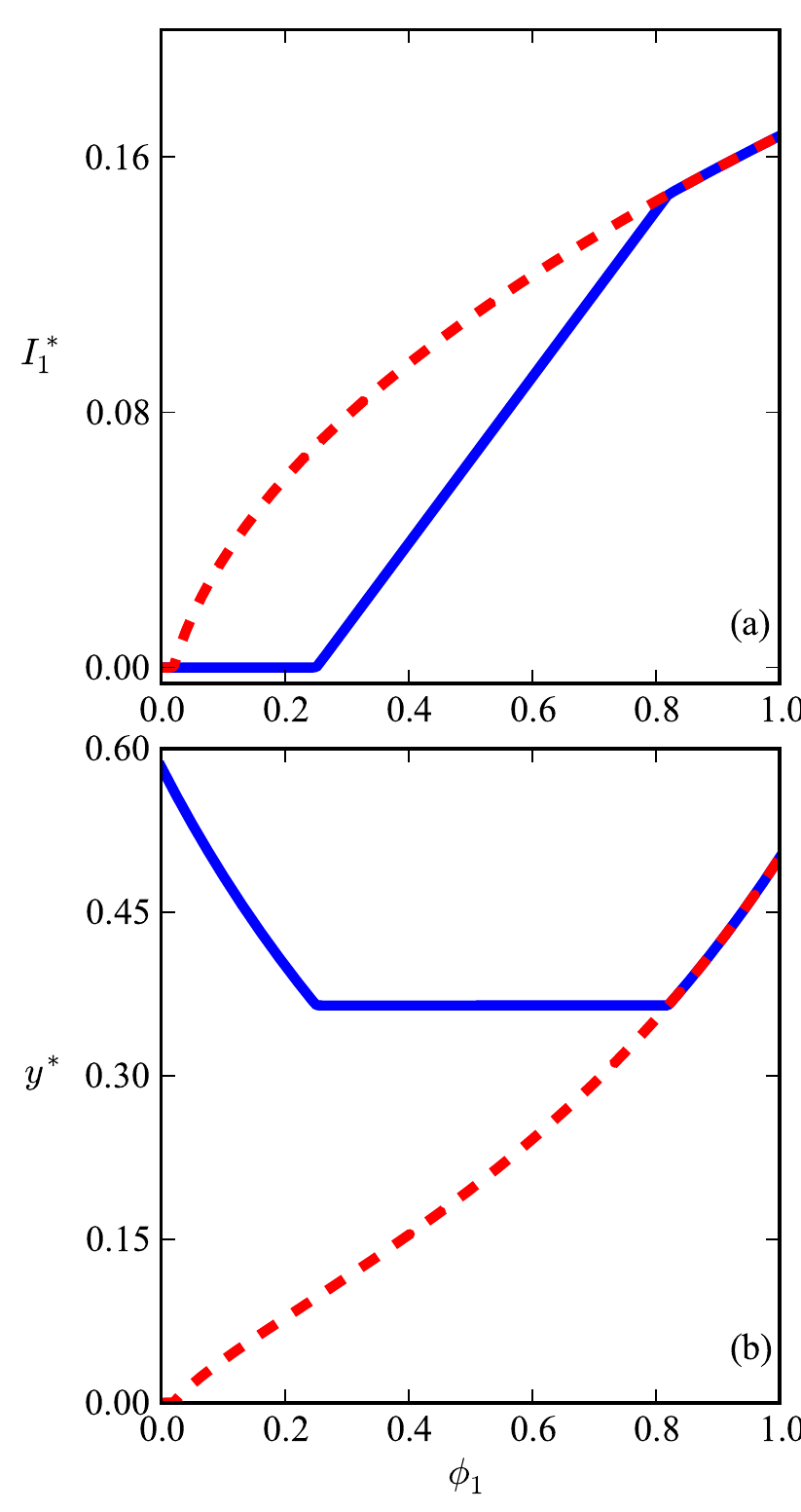}} \caption{
    The effect of competition between the two pathogens $P_1$ and
    $P_2$ on the amount of disease. (a) $I_1^*$ is plotted as a
    function of $\phi_1=K_1/(K_1+K_2)$ at $\beta_{22}=\beta_{21}=0$ (red, dashed curve), in which
    case $P_2$ is absent and no competition occurs; and at $\beta_{22}=8$,
  $\beta_{21}=0.9$ (blue, solid curve), when $P_2$ is present and competes with $P_1$. (b) the
disease severity $y^*$ is plotted versus $\phi_1$ at $\beta_{22}=\beta_{21}=0$ (red,
dashed curve) and at $\beta_{22}=8$, $\beta_{21}=0.9$ (blue, solid
curve). Other parameter values are the same as in \fig{fig:pd-freqres-vs-a1-somespec}.}
\label{fig:i1-i2-comp-vs-a1}
\end{figure}

When the pathogen-host specialization is not complete
($\beta_{21}>0,\,\beta_{12}>0$) and the two pathogens coexist (at intermediate
values of $\phi_1$), the two pathogens compete with each other for host
tissue. The effect of this competition on the amount of disease is
illustrated in \fig{fig:i1-i2-comp-vs-a1}: panel (a) shows the amount
of host tissue $I_1^*$ infected by $P_1$ versus $\phi_1$, when $P_2$ is
present (blue, solid curve) and absent (red, dashed curve). The
presence of $P_2$ decreases $I_1^*$ across almost the whole range of
$\phi_1$, except for the largest values, where $P_1$ dominates the
population and $P_2$ dies out. However, if we look at the overall
disease severity caused by both pathogens
[\fig{fig:i1-i2-comp-vs-a1}(b)], the presence of $P_2$ makes it
larger. Therefore, although the application of $P_2$ as a biocontrol
measure suppresses $P_1$, it also increases the overall disease
severity. This could only be considered as a reasonable control
measure, if pathogen $P_1$ is much less desirable (e.g. it produces a
mycotoxin that is not produced by $P_2$ or has a higher risk of
developing fungicide resistance compared to $P_2$), while a certain degree of infection
with $P_2$ can still be tolerated. 

A prominent example where this model can be applied in the context of
biocontrol is illustrated with \emph{Aspergillus flavus}, a fungal
pathogen that can infect a variety of crops, including maize,
cottonseed, peanuts, and tree nuts. Some strains of \emph{A. flavus}
produce aflatoxins, toxic carcinogenic fungal metabolites that
contaminate food (aflatoxigenic strains), while others do not
(atoxigenic strains). Atoxigenic strains are deliberately applied to
crops as a form of biocontrol to mitigate aflatoxin contamination
\citep{meja+12}. 
In the example case considered in \fig{fig:i1-i2-comp-vs-a1}, when the
atoxigenic strain $P_2$ is applied, it may outcompete the
aflatoxigenic strain $P_1$, leading to its eradiction. Our model
predicts that this occurs at a low enough proportion of the host
variety $H_1$ in the mixture (i.\,e. the cropping ratio $\phi_1$
should be below a certain value marked by a vertical dotted line in
\fig{fig:i1-i2-comp-vs-a1}). 

Thus, mixing host varieties reduces the overall disease severity if
each of the pathogens performs better on its preferred host. In this
case, an optimal proportion of host varieties in the mixture lies in
the intermediate range, over which the two pathogens exhibit stable
co-existence. This result is in agreement with previous theoretical
studies \citep{li10} and also explains some experimental findings
\citep{zhmc13}.

In addition, within the range of maximal overall suppression 
of disease, the ratio of the two pathogens can be
controlled by varying the proportion of hosts in the mixture
[\fig{fig:pd-freqres-vs-a1-somespec}(a) and (b)]. This can be useful,
if one of the pathogens is much less desirable, for example, because
of mycotoxin production or the risk of fungicide resistance. 

\subsection{What is the optimal number of components to use in a host mixture?}
\label{sec:optncomp}

\begin{figure}
  \centerline{\includegraphics[width=0.8\textwidth]{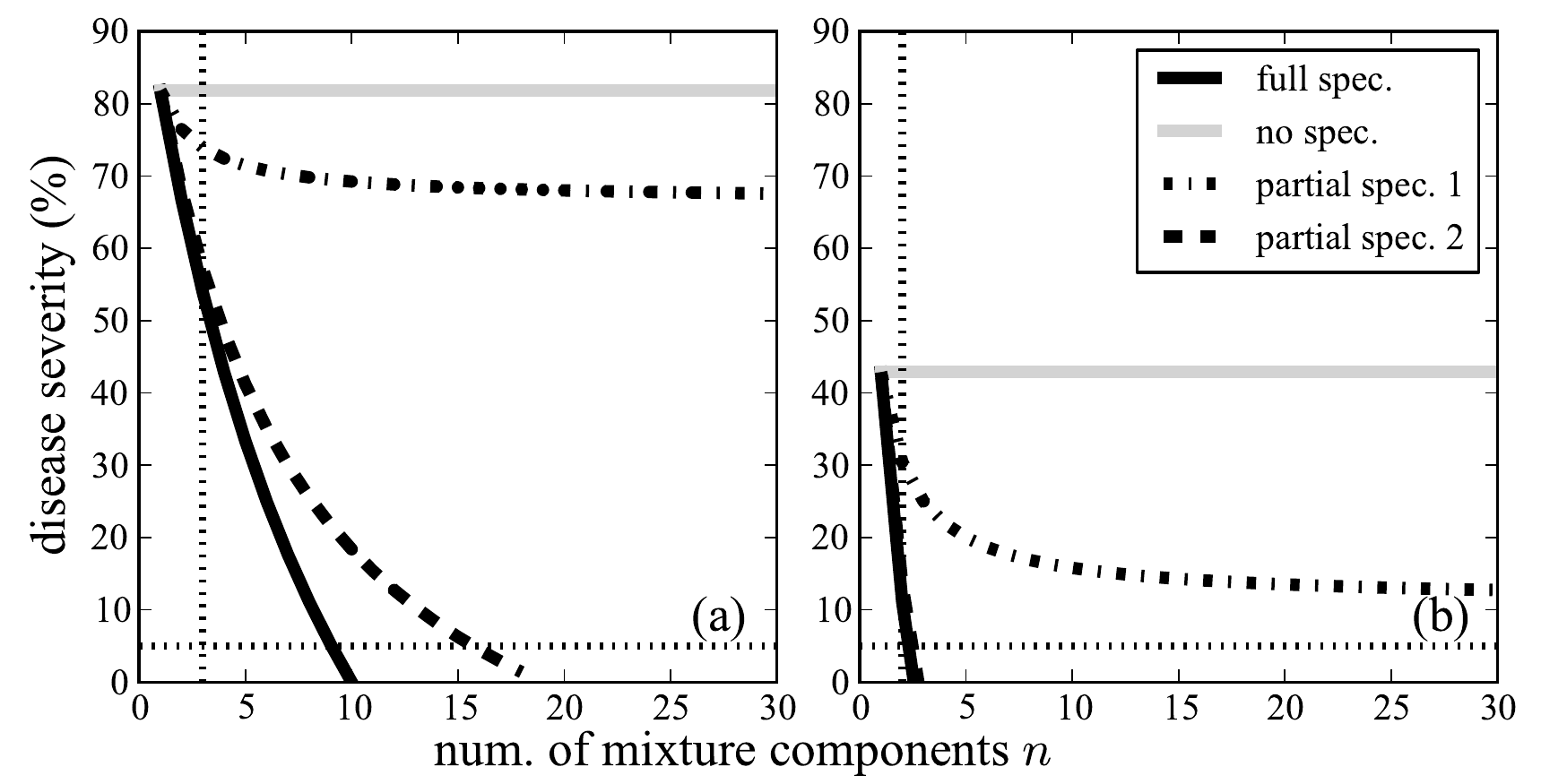}} \caption{
    Disease severity at the infected equilibrium versus the number of
    components in the host mixture plotted according to
    \eq{eqap:sevtot-ps} in the case of no specialization (grey solid),
    full specialization (solid), partial specialization with the
    specialization index $\sigma= \beta_\mrm{nd}/\beta_\mrm{d}=0.5$
    (dash-dotted) and $\sigma=0.05$ (dashed). Paramer values: (a)
    pathogen with high transmission $\beta_\mrm{d} = 2$; (b) pathogen
    with low transmission $\beta_\mrm{d} = 0.5$. The rest of
    parameters are the same in (a) and (b): $K_\mrm{tot}=1$,
    $\mu=0.2$, $r=0.1$. Dotted horizontal curve shows an example of a
    maximum disease severity, $S_\mrm{acc}= 5\,\%$, that is still
    economically acceptable. Dotted vertical lines show the optimal
    number of components $n_\mrm{opt2}=3$ [panel (a)] and
    $n_\mrm{opt2}=2$ [panel (b)], according to \eq{eq:ncomp-opt2}
    taking $\Delta S = 10$\,\%, for the dashed curves.}
\label{fig:dis-sev-vs-ncomp}
\end{figure}

So far we considered a host mixture with only two components. Does
adding more components to the mixture lead to better disease control? 
We use the mathematical framework of
Eqs.\,(\ref{eq:hi-nhost-npath-ps})-(\ref{eq:ii-nhost-npath-ps}) to
address this question. In the simplified case of partial
specialization all elements of the transmission matrix $\vec{B}$ are
positive. All the diagonal elements are equal to $\beta_\mrm{d}$ and the
non-diagonal ones are equal to $\beta_\mrm{nd}$, with $\beta_\mrm{d} > \beta_\mrm{nd}>0$ (see
\eq{eq:bmatr-ps}). 
In this case, we determined the analytical expression for the total
disease severity at the infected equilibrium 
\begin{equation}\label{eqap:sevtot-ps}
y_\mrm{tot}^*(n) = r \frac{ \left( \beta_\mrm{d} + (n-1) \beta_\mrm{nd}
    \right) K_\mrm{tot} - \mu n } {n  \mu (\mu - r) + r K_\mrm{tot} \left( \beta_\mrm{d} + (n-1) \beta_\mrm{nd} \right)}.
\end{equation}
%
Using this expression, we plotted in \fig{fig:dis-sev-vs-ncomp} the
disease severity as a function of the number of components in the
mixture $n$. Panel (a) illustrates the case of a pathogen with the
high rate of transmission and panel (b) shows the case a pathogen with
the intermediate rate of transmission. The grey solid curves represent
the homogeneous case when $\beta_\mrm{nd} = \beta_\mrm{d}>0$,
i.\,e. no specialization, every pathogen strain or species is equally
likely to infect every host. Evidently, in this case the disease
severity is independent of the number of mixture components.
%
%
In all other cases considered in \fig{fig:dis-sev-vs-ncomp}, the
disease severity decreases with $n$. The black solid curves in
\fig{fig:dis-sev-vs-ncomp} illustrate the case of full
specialization, when $\beta_\mrm{nd}=0,\,\beta_\mrm{d}>0$. In this case, the
disease severity decreases steeply with increasing $n$,
eventually reaching zero. The dashed curves in
\fig{fig:dis-sev-vs-ncomp} correspond to intermediate cases with
different degrees of partial specialization. As the degree of host
specialization increases, the decrease in disease severity becomes
stronger.

Can one eradicate the disease by adding a large enough number of
components to the host mixture?
%
%
As we increase the number of components in the host mixture, each
pathogen strain can infect less of its preferred host. At the limit of
very large $n$, the amount of preferred host tissue available for each pathogen
strain is so small that they are not able to survive only on it.
Therefore, whether we can eradicate the disease depends on the ability
of pathogen strains to survive on hosts that are not their
favorite. This is determined by the parameter $R_\mrm{0nd} =
\beta_\mrm{nd} K_\mrm{tot} / \mu$, which is the basic reproductive
number of pathogen strains as a whole in the absence of their
preferred hosts.  If $R_\mrm{0nd}>1$, then pathogen strains can
survive in the absence of their preferred hosts. In this case, disease
severity tends to a constant positive value at large $n$ and never
decreases to zero (dash-dotted curve in
\fig{fig:dis-sev-vs-ncomp}). In contrast, when $R_\mrm{0nd}<1$,
pathogen strains die out in the absence of their preferred hosts.



We take the the limit of very large $n$ in \eq{eqap:sevtot-ps} and
find that the disease severity is proportional to $R_\mrm{0nd}-1$ in
this case:
%
%
\begin{equation}\label{eq:sevtot-ps-limn}
y_\mrm{tot}^*(n)_{n \to \infty} = r \frac{ R_\mrm{0nd} - 1 } { \mu  + r (R_\mrm{0nd} - 1)},
\end{equation}
where 
\be\label{eq:brn-nd}
R_\mrm{0nd} = \beta_\mrm{nd} K_\mrm{tot} / \mu
\ee
is the basic reproductive number of pathogen strains overall in the
absence of their preferred hosts. It follows from
\eq{eq:sevtot-ps-limn} that if $R_\mrm{0nd} \leq 1$, then the disease
severity will eventually reach (or approach) zero as we increase $n$.
However when $R_\mrm{0nd}>1$, the disease severity will approach a
constant positive value given by \eq{eq:sevtot-ps-limn}.
This means that, by increasing the number of components in the
mixture, we decrease (eventually to zero) the impact of host-specialized infections
characterized by rate $\beta_\mrm{d}$. However, the impact of non-specialized
infections characterized by $\beta_\mrm{nd}$ remains unchanged with the
corresponding severity given by \eq{eq:sevtot-ps-limn}.


From the expression for the disease severity in \eq{eqap:sevtot-ps}, one
can determine the optimal number of components to use in the mixture. One way
to do this is to define an economically acceptable disease severity,
$y_\mrm{acc}$, (for example 5\,\%), and then determine the number of
components in the mixture that decrease the disease severity down to
$y_\mrm{acc}$. This is done by solving \eq{eqap:sevtot-ps} with respect
to $n$. As a result, we obtain 
\be\label{eq:ncomp-opt}
n_\mrm{opt1} = r K_\mrm{tot} \frac{(\beta_d - \beta_\mrm{nd}) (1 - y_\mrm{acc}) } {\mu (r + y_\mrm{acc} (\mu - r)) - r \beta_\mrm{nd} K_\mrm{tot} ( 1 - y_\mrm{acc})}.
\ee
%
%
Here, $n_\mrm{opt1}$ is the number of mixture components at which the disease
severity $y_\mrm{acc}$ is reached.  This is illustrated in
\fig{fig:dis-sev-vs-ncomp}, where the horizontal dashed line
corresponds to $y_\mrm{acc}= 5\,\%$. The values of $n$ at which
this line intersects with disease severity curves correspond to
optimum $n_\mrm{opt1}$ given by \eq{eq:ncomp-opt}. The optimum shifts
to larger values with decreasing degrees of specialization
[e.\,g. from $n_\mrm{opt1}=9$ for the solid curve corresponding to full
specialization to $n_\mrm{opt1}=16$ for the dashed curve representing
partial specialization in \fig{fig:dis-sev-vs-ncomp}(a)].



Another way to determine an optimal number of mixture components uses
the fact that $y^*(n)$ decreases with $n$, but also considers that the rate of this
decrease (i.\,e. the derivative $\frac{d y^*(n)} {d n}$) decreases
with $n$. Hence, the benefit of adding one more component to a
mixture that already has $n$ components decreases with increasing $n$.
Because of this, the dependence $y^*(n)$ eventually saturates to a
constant value given by \eq{eq:sevtot-ps-limn}.
%
 %
Therefore, one can define a minimum decrease in disease severity due to
adding one more host variety to the mixture $\Delta y_\mrm{min}$ that
is still economically plausible. The number of mixture components at
this minimum is optimal, i.\,e. $n=n_\mrm{opt2}$.
%
%
Mathematically, $n_\mrm{opt2}$ can be found from the equation
$y_\mrm{tot}^*(n_\mrm{opt2}-1) - y_\mrm{tot}^*(n_\mrm{opt2}) = \Delta
y_\mrm{min}$, where $y_\mrm{tot}^*(n)$ is given by
\eq{eqap:sevtot-ps}. The solution reads as
\be\label{eq:ncomp-opt2}
n_\mrm{opt2} = \frac{ \sqrt{\Delta y} \left[ \mu^2 - r \left(
      K_\mrm{tot} (2 b_d - 3 b_\mrm{nd}) + \mu \right) \right] + \sqrt{4
    (b_d - b_\mrm{nd}) r K_\mrm{tot} \mu^2 + \Delta y C }} {2 \sqrt{\Delta
  S} C^2},
\ee
where
\be
C = \mu^2 + r (b_\mrm{nd} K_\mrm{tot} - \mu).
\ee
This is also illustrated in \fig{fig:dis-sev-vs-ncomp}, where the
dotted vertical lines shows $n_\mrm{opt2}=3$ [panel (a)] and
$n_\mrm{opt2}=2$ [panel (b)] that correspond to the
severity curves for the case of strong partial specialization (dashed curves). When
the degree of specialization is increased further up to full
specialization (solid curve), $n_\mrm{opt2}$ shifts to the larger value of
four.

We expect mixtures to be more effective against pathogens with
intermediate and low transmission [cf. panels (a) and (b) in
\fig{fig:dis-sev-vs-ncomp}]. In \fig{fig:dis-sev-vs-ncomp}(b) a
mixture with three components not only decreased the disease below the
acceptable level [optimum number of components, according to
\eq{eq:ncomp-opt}], but even eradicated the pathogen. A two-component
mixture provided an economical optimum, according
\eq{eq:ncomp-opt2}. In contrast, for pathogens with high
transmission [\fig{fig:dis-sev-vs-ncomp}(a)], mixtures with more
components need to be used to reach the optimal effects.


The optimum number of components in the mixture, defined according to
\eq{eq:ncomp-opt}, can only be found if the acceptable severity
$y_\mrm{acc}$ can be reached by increasing $n$ (that is when
$R_\mrm{0nd}<1$). This restriction is removed in the definition based on
\eq{eq:ncomp-opt2}. But even in cases when $y_\mrm{acc}$ can be
reached by increasing $n$, the second definition seems to be more
plausible, since it incorporates the economic costs of introducing an
additional component into the mixture. However, it does not ensure
that the disease will be reduced down to an acceptable value. Hence,
additional disease control measures (e.\,g. applications of
fungicides) may need to be implemented in order to further reduce the
disease.








%

\section{Discussion}
\label{sec:discussion}


We have shown that when a population of crop plants is exposed to two
host-specialized pathogen strains or species, the overall severity of
both diseases is smaller in the mixture of two host varieties than in
either of the pure stands. We obtained analytical expressions for the
disease reduction which allowed us to quantify it across the whole range
of parameters. These findings may help to identify crop cultivars to be deployed in 
mixtures that will successfully control diseases prevalent in a given
region.
%
The overall disease severity can be minimized over a range of mixing
ratios. The two pathogens coexist in this range and further adjusting
the mixing ratio within this range makes it possible to control the relative
abundance of each pathogen. This can be useful when one of the
pathogens is less desirable, for example due to mycotoxin production
or fungicide resistance, while a certain amount of the other pathogen
can be tolerated. Alternatively, the mixing ratio can be adjusted
within this optimal range to increase the economic output of the crop,
if the two host varieties differ in their quality or commercial value.

We also generalized the model to describe host mixtures with more than
two components. We find that when there is a degree of host
specialization, the overall disease severity decreases with the number
of components in the mixture. The more specialized the host-pathogen
pairs are, the stronger is the decrease in the disease severity. Based
on this understanding, we proposed ways to determine economically
optimal numbers of components in host mixtures. Furthermore, this more
general framework is capable of describing many hosts exposed to many
pathogen strains or species and can also be used to better understand
plant-pathogen dynamics in natural ecosystems, such as \emph{Linum
marginale}--\emph{Melampsora lini} \citep{thbu+02}, or \emph{Plantago
lanceolata}--\emph{Podosphaera plantaginis} \citep{la07}. Local adaptation was
detected in these natural interactions \citep{thbu+02,la07}, hence the insight we gained in
the case of partial specialization may advance our understanding of
evolutionary forces operating in these wild plant-pathogen systems.

Four distinct mechanisms of disease reduction by host mixtures are
described in the literature \citep{chwo84,wo85,figa+00}: (i) the
effect of reduced density of susceptibles; (ii) the ``barrier
effect''; (iii) induced resistance; and (iv) competition between
pathogens.
%
In scenario (i) the disease is reduced in the mixture simply because it
has less of the susceptible variety than the susceptible pure stand.
This ``reduced density'' effect can be observed most
clearly by comparing the amount of disease in two pure stands of the
susceptible variety, which differ only in planting density
\citep{chwo84}.
The introduction of the resistant variety further reduces the disease
in the mixture (scenario (ii)), because the transmission between
susceptible hosts is hindered (a resistant ``barrier'' is created
between adjacent susceptible plants). Induced resistance (scenario
(iii)) takes place when spores of an avirulent pathogen activate a
host resistance mechanism that is also effective against another
pathogen (or another race of the same pathogen), which is normally
able to infect the host \citep{chwo84,lava+95,lahu+05}. Finally, in
scenario (iv) mixing host cultivars is expected to make the pathogens
compete with each other for host tissue \citep{figa+00,ohsa06}.

The ``reduced density'' effect originally referred to the mixture of a
susceptible and a resistant variety \citep{chwo84}. Hence, it cannot
lead to a disease level lower than in the pure stand of the resistant
variety. Here we extended the notion of the ``reduced density'' effect
to the case of two or more host-specialized pathogen strains or
species. For example, this may correspond to host 1 being susceptible
to pathogen 1, but resistant to pathogen 2 and host 2 being
susceptible to pathogen 2, but resistant to pathogen 1. We find that
it is only in such cases that disease level in the mixture is lower than in
both pure stands.

Our results also indicate that although the presence of the second
pathogen may suppress the population of the first one, the overall
disease severity due to both of them increases
(\fig{fig:i1-i2-comp-vs-a1}). Hence, according to our model, the
competition between pathogens (scenario (iv)) alters the relative
abundunce of each of them, but is not capable of reducing the overall
disease severity.

Our model does not include the ``barrier'' effect, since it does not
explicitly consider the spatial dependence of pathogen dispersal (see
\sec{sec:model}). Also, induced resistance
\citep{chwo84,lava+95,lahu+05} was not
considered. Therefore, we likely underestimate the effect of host mixtures
on disease reduction. However, the model can be readily extended to include both
of these effects.
In this way, a unified mathematical framework for description of the
effect of host mixtures on plant disease can be developed on the basis
of the model presented here. This would allow one to better understand
the relative contributions of each of these effects in disease
reduction and design better host mixtures.

Here we focused on the benefit host mixtures may provide in terms of
reduction of disease. But how do host mixtures influence pathogen evolution over 
longer time scales?  
To investigate the durability of resistance in the context considered
here, when a mixture of two host varieties is exposed to two different
pathogens (or different races of the same pathogen), each specialized
on one of the host varieties, one needs to first consider the
likelihood of emergence of a third pathogen race, which can infect
both host varieties equally successfully (the ``super-race''), but
might bear a fitness cost associated with the extended host range. In
this scenario, the conditions under which this pathogen race will be
favored by selection need to be considered. This can be done by using
mathematical frameworks similar to the ones developed in
\citep{iabo+12,bogi03} and is a promising direction for future study.

\section{Acknowledgements}
\label{sec:acknowledg}
AM and SB gratefully acknowledge support by the ERC advanced grant
PBDR 268540. AM would like to thank Gabriel Leventhal for helpful discussions.

%

\appendix
\numberwithin{equation}{section}

\section{Electronic Supplementary Material}

\subsection{Fixed points of the two hosts--two pathogens model}
\label{apsec:fps}


The system of equations (\ref{eq:2host2fung-1})-(\ref{eq:2host2fung-4})
has six fixed points: 

FP1: $H ^*_1=K_1,\, H ^*_2=K_2,\, I ^*_1=I ^*_2=0$; both pathogens die out

FP2: $H ^*_1 \ne 0,\, H^*_2 \ne 0,\, I_1 \ne 0,\, I^*_2=0$; $P_1$ wins

FP3 $H^*_1 \ne 0,\, H^*_2 \ne 0,\, I^*_1 \ne 0,\, I^*_2=0$; $P_1$ wins

FP4 $H^*_1 \ne 0,\, H^*_2 \ne 0,\, I^*_1 \ne 0,\, I^*_2 \ne 0$; $P_1,\, P_2$ coexist

FP5 $H^*_1 \ne 0,\, H^*_2 \ne 0,\, I^*_1 = 0,\, I^*_2 \ne 0$; $P_2$ wins

FP6 $H^*_1 \ne 0,\, H^*_2 \ne 0,\, I^*_1 = 0,\, I^*_2 \ne 0$; $P_2$ wins


Over the whole range of biologically plausible parameter values ($r>0$;
$K_1>0$; $K_2>0$; $\beta_{ij} \geq 0$, $i,j=1..2$; $\mu>0$ ), only
four of the fixed points (we denote them as FP1, FP2, FP4, FP5) correspond to positive values of the
corresponding amounts of host tissue $H^*_1$, $H^*_2$, $I^*_1$,
$I^*_2$, while the fixed points FP3 and FP6 correspond to at least one
of the quantities $H^*_1$, $H^*_2$, $I^*_1$, $I^*_2$ being negative,
which is biologically unrealistic.
We are interested in the case when at least one of the pathogens
survives. Therefore, we consider the expressions for $H^*_1$, $H^*_2$,
$I^*_1$, $I^*_2$ at the fixed points FP2, FP4, FP5.

Values of $I_1$, $I_2$, $H_{1}$ and $H_{2}$ at the fixed point FP2 are given by
\begin{equation}\label{eq:i1-fp2}
I_{1\mrm{\{FP2\}}} = r \frac{ \beta_{11} \beta_{12} K - \mu_1 (\beta_{11} + \beta_{12}) +
  \sqrt{B_\mrm{FP2}} } {2 \beta_{11} \beta_{12} \mu_1}, \: I_{2\mrm{\{FP2\}}}=0,
\end{equation}
\begin{equation}\label{eq:h1-fp2}
H_{1\mrm{\{FP2\}}} =  \frac{ - \beta_{11} \beta_{12} K + \mu_1 (\beta_{11} - \beta_{12}) +
  \sqrt{B_\mrm{FP2}} } {2 \beta_{11} (\beta_{11} - \beta_{12} )},
\end{equation}
\begin{equation}\label{eq:h1-fp2}
H_{2\mrm{\{FP2\}}} =  \frac{  \beta_{11} \beta_{12} K + \mu_1 (\beta_{11} - \beta_{12}) -
  \sqrt{B_\mrm{FP2}} } {2 \beta_{12} (\beta_{11} - \beta_{12} )},
\end{equation}
where
\begin{equation}\label{eq:b-fp2}
B_\mrm{FP2} = 4 \beta_{11} \beta_{12} (\beta_{11} - \beta_{12}) \mu_1 \phi_1 K + \left[
  \beta_{11} \beta_{12} K -  (\beta_{11} - \beta_{12}) \mu_1 \right]^2.
\end{equation}

Values of $I_1$, $I_2$, $H_{1}$ and $H_{2}$ at the fixed point FP5 are given by
\begin{equation}\label{eq:i2-fp5}
I_{2\{\mrm{FP5}\}} = r \frac{ \beta_{22} \beta_{21} K - \mu_2 (\beta_{22} + \beta_{21}) +
  \sqrt{B_\mrm{FP5}} } {2 \beta_{22} \beta_{21} \mu_2}, \: I_{1\mrm{\{FP5\}}}=0,
\end{equation}
\begin{equation}\label{eq:h1-fp5}
H_{1\mrm{\{FP5\}}} =  \frac{ \beta_{22} \beta_{21} K + \mu_2 (\beta_{22} - \beta_{21}) -
  \sqrt{B_\mrm{FP5}} } {2 \beta_{21} (\beta_{22} - \beta_{21} )},
\end{equation}
\begin{equation}\label{eq:h2-fp5}
H_{2\mrm{\{FP5\}}} =  \frac{ - \beta_{22} \beta_{21} K + \mu_2 (\beta_{22} - \beta_{21}) +
  \sqrt{B_\mrm{FP5}} } {2 \beta_{22} (\beta_{22} - \beta_{21} )},
\end{equation}
where 
\begin{equation}\label{eq:b-fp5}
  B_\mrm{FP5} = -4 \mu_2 \phi_1 K \beta_{22} \beta_{21} (\beta_{22} - \beta_{21}) +
 \left[ \beta_{22} \beta_{21} K + \mu_2 (\beta_{22} - \beta_{21}) \right]^2.
\end{equation}

Values of $I_1$, $I_2$, $H_{1}$ and $H_{2}$ at the fixed point FP4 are given by
\begin{equation}\label{eq:i1-fp4}
I_{1\mrm{\{FP4\}}} = r \frac{ \beta_{22} \beta_{21} \mu_1 \left( C_{21} \mu_1 - K C_{-} \right) + \left[ (\phi_1 C_{-} + \beta_{12} \beta_{21}) C_{-}
      K - C_{21} C_{+} \mu_1 \right]
 \mu_2 + \beta_{12} \beta_{21}
  C_{21} \mu_2^2 } {C_{-} (\beta_{21} \mu_1 - \beta_{11}
\mu_2) (\beta_{12} \mu_2 - \beta_{22} \mu_1)},
\end{equation}
\begin{equation}\label{eq:i2-fp4}
I_{2\mrm{\{FP4\}}} = r \frac{ (\beta_{22} \mu_1 - \beta_{12} \mu_2) \left[
    \beta_{11} C_{-} K - C_{12} (\beta_{11} \mu_2 - \beta_{21} \mu_1) \right] -
  C_{-}^2 \phi_1 K \mu_1 } { C_{-} (\beta_{11} \mu_2 - \beta_{21} \mu_1) (\beta_{22} \mu_1 - \beta_{12} \mu_2) },
\end{equation}
\begin{equation}\label{eq:h1-fp4}
H_{1\mrm{\{FP4\}}} = \frac{\beta_{22} \mu_1 - \beta_{12} \mu_2} {\beta_{11} \beta_{22}
- \beta_{12} \beta_{21}},
\end{equation}
\begin{equation}\label{eq:h2-fp4}
H_{2\mrm{\{FP4\}}} = \frac{\beta_{11} \mu_2 - \beta_{21} \mu_1} {\beta_{11} \beta_{22}
- \beta_{12} \beta_{21}},
\end{equation}
where 
\begin{equation}\label{eq:c}
C_{12} = \beta_{11} - \beta_{12}, \: C_{21} = \beta_{22} - \beta_{21}
\end{equation}
\begin{equation}\label{eq:fp4-cmin}
C_{-} =  \beta_{11} \beta_{22} - \beta_{12} \beta_{21},
\end{equation}
\begin{equation}\label{eq:fp4-cplus}
C_{+} =  \beta_{11} \beta_{22} + \beta_{12} \beta_{21}.
\end{equation}
%
Next, we can obtain the expressions for the disease severity
\begin{equation}\label{eq:pdis-eq-app}
y^* = \frac{I_1^* + I_2^*} {I_1^* + I_2^* + H_1^* + H_2^*}
\end{equation}
and for the frequency of pathogen 2 at equilibrium
\begin{equation}\label{eq:f2-def}
f^*_2 = \frac{I_2^* } {I_1^* + I_2^*}
\end{equation}
corresponding to different fixed points described above. Since we
consider only two pathogens, the frequency of the pathogen 1 at
equilibrium is then $f^*_1 = 1 - f^*_2$.

In order to determine the dependence of $y^*$ and $f^*_2$ on the
proportion of host variety 1 in the mixture $\phi_1$ and the fungicide
concentration $C$, we choose from the expressions
(\ref{eq:i1-fp2})-(\ref{eq:h1-fp4}) those which correspond to a
stable fixed point at a given value of $\phi_1$ and $C$, and substitute
them in \eq{eq:pdis-eq-app} and \eq{eq:f2-def}. The result of this procedure is shown in
\fig{fig:pd-freqres-vs-a1-somespec}.



\subsection{Linear stability of the fixed points and invasion thresholds for the two hosts--two pathogens model}
\label{apsec:2h2p-linstab}

The trivial fixed point FP1 (both
pathogens die out) is realized when both $R_{01}<1$ and $R_{02}<1$,
where 
\begin{equation}
R_{01} = (\beta_{11} K_1 + \beta_{12} K_2)/\mu,
\end{equation}
\begin{equation}
R_{02} = (\beta_{21} K_1 + \beta_{22} K_2)/\mu,
\end{equation}
That is when $\beta_{11}<\beta_\mrm{11b}$ and $\beta_{22}<\beta_\mrm{22b}$, where
\begin{equation}
\beta_\mrm{11b} = ( \mu  - K_2 \beta_{12})/ K_1,
\end{equation}
\begin{equation}
\beta_\mrm{22b} = ( \mu  - K_1 \beta_{21})/ K_2.
\end{equation}
If $\beta_{11}> \beta_\mrm{11b}$, but $\beta_{22}< \beta_\mrm{22b}$, then the pathogen $P_1$
wins. And vice versa, if $\beta_{22}> \beta_\mrm{22b}$, but $\beta_{11}< \beta_\mrm{11b}$,
then the pathogen $P_2$ wins.

Now, we determine the threshold for invasion of the pathogen $P_2$,
when the host population is already infected at equilibrium with the
pathogen 1 (this corresponds to the fixed point FP2). We linearize the
system (\ref{eq:2host2fung-1})-(\ref{eq:2host2fung-4}) in the
vicinity of the fixed point FP2 and search for the conditions under
which it becomes unstable with respect to the invasion of $P_2$. In
order to do this, we only need to consider the linearized equation for
$I_2$, since it becomes uncoupled from the other equations:
\begin{equation}\label{eq:i2-fp2-lin}
\frac{d I_2}{d t} = \lambda_2 I_2,
\end{equation}
where
\begin{equation}\label{eq:lambda2-fp2-lin}
\lambda_2 = \beta_{21} H_{1\{FP2\}}^* + \beta_{22} H_{2\{FP2\}}^* - \mu
\end{equation}
is the growth rate. Then, we obtain the corresponding basic
reproductive number 
%
\begin{equation}\label{eq:r02-fp2}
R_{02\{FP2\}} = \left( \beta_{21} H_{1\{FP2\}}^* + \beta_{22} H_{2\{FP2\}}^* \right)/ \mu.
\end{equation}
Here, the equilibrium values of the susceptible host density
$H_{1\{FP2\}}^*$ and $H_{2\{FP2\}}^*$ are given by \eq{eq:h1-fp2} and \eq{eq:h1-fp2}.

From \eq{eq:r02-fp2} we obtain threshold value of the transmission
rate, above which the pathogen 2 can invade:
%

\begin{equation}\label{eqap:b22c}
b_\mrm{22c} = \frac{ \beta_{12} \left( \beta_{12} \beta_{21} K + \left[ 2(\beta_{11}
        - \beta_{12})  - \beta_{21}+ \beta_{12} \beta_{21}/\beta_{11} \right] \mu - \beta_{21}/\beta_{11} \sqrt{B_\mrm{FP2}} \right) }{ (\beta_{11} - \beta_{12}) \mu + \beta_{11} \beta_{12} K - \sqrt{B_\mrm{FP2}}  },
\end{equation}
where $B_\mrm{FP2}$ is given by \eq{eq:b-fp2}.

%

Similarly, we determine the threshold for invasion of pathogen 1
($I_1$), when the host population is already infected at equilibrium
with pathogen 2 (this corresponds to the fixed point FP5). Again, we
linearize the system
(\ref{eq:2host2fung-1})-(\ref{eq:2host2fung-4}) in the vicinity
of the fixed point FP5 and search for the conditions under which it
becomes unstable with respect to the invasion of pathogen 1. In
order to do this, we only need to consider the linearized equation for
$I_1$, since it becomes uncoupled from other equations:
\begin{equation}\label{eq:i1-fp5-lin}
\frac{d I_1}{d t} = \lambda_1 I_1,
\end{equation}
where
\begin{equation}\label{eq:lambda2-fp2-lin}
\lambda_1 = \beta_{11} H_{1\{FP5\}}^* + \beta_{12} H_{2\{FP5\}}^* - \mu
\end{equation}
is the growth rate. Then, we obtain the corresponding basic
reproductive number 
%
\begin{equation}\label{eq:r02-fp2}
R_{01\{FP5\}} = \left( \beta_{11} H_{1\{FP5\}}^* + \beta_{12} H_{2\{FP5\}}^* \right)/ \mu.
\end{equation}
Here, the equilibrium values of the susceptible host density
$H_{1\{FP5\}}^*$ and $H_{2\{FP5\}}^*$ are given by \eq{eq:h1-fp5} and \eq{eq:h2-fp5}.

Similarly the critical value of the infection rate $b_\mrm{11c}$,
above which the pathogen 2 can invade, reads
\begin{equation}\label{eqap:b11c}
b_\mrm{11c} = \frac{ \beta_{21} \left( \beta_{21} \beta_{12} K + \left[ 2(\beta_{22}
        - \beta_{21})  - \beta_{12}+ \beta_{21} \beta_{12}/\beta_{22} \right] \mu - \beta_{12}/\beta_{22} \sqrt{B_\mrm{FP5}} \right) }{ (\beta_{22} - \beta_{21}) \mu + \beta_{22} \beta_{21} K - \sqrt{B_\mrm{FP5}}  },
\end{equation}
where $B_\mrm{FP5}$ is given by \eq{eq:b-fp5}.
%

\subsection{Disease severity when $n$ hosts are exposed to $n$ pathogens}
\label{secap:dissev-nh-np}

Consider the case when $n$ hosts are exposed to $n$ pathogens and the
transmission matrix has a simple form given by \eq{eq:bmatr-ps}, where
every diagonal element of the matrix $\vec{B}$ is equal to $\beta_\mrm{d}$
and every non-diagonal element is $\beta_\mrm{nd}$. We assume that every
host variety is planted at the same proportion, i.\,e. $K_i = K$. In
addition, we assume that all healthy and infected hosts start with
the same initial conditions. Hence, their dynamics are the same and is
described by a simplified system of
Eqs.\,(\ref{eq:hi-nhost-npath-ps})-(\ref{eq:ii-nhost-npath-ps})
\begin{linenomath}
\begin{align}
\frac{d H_p}{d t} &= r ( K - H_p) - \beta_\mrm{eff} I_p H_p, \label{eqap:hi-nhost-npath-ps}\\ 
\frac{d I_p}{d t} & =  \beta_\mrm{eff} I_p H_p - \mu  I_p, \label{eqap:ii-nhost-npath-ps}
\end{align}
\end{linenomath}
where $\beta_\mrm{eff} = \beta_\mrm{d} + (n-1) \beta_\mrm{nd}$. Then, the values
of host densities at the infected equilibrium read:
\begin{equation}\label{eqap:hp-ip-ps}
H_p^* = \frac{\mu}{\beta_\mrm{eff}} , \: I_p^* = \frac{r K}
{\mu} \left( 1 - \frac{ \mu} {\beta_\mrm{eff} K } \right).
\end{equation}
The total amounts of healthy and infected hosts are again obtained by
multiplying $H_p$ and $I_p$ by $n$:
\begin{linenomath}
\begin{align}\label{eqap:htot-itot-ps}
H_\mrm{tot}^* = n H_p^* =  \frac{n \mu}{\beta_\mrm{d} + (n-1) \beta_\mrm{nd}} , \:\\ I_\mrm{tot}^* = n I_p^*  = \frac{r K_\mrm{tot}}
{\mu} \left( 1 - \frac{ n \mu} {\left( \beta_\mrm{d} + (n-1) \beta_\mrm{nd} \right) K_\mrm{tot}} \right).
\end{align}
\end{linenomath}
In this case, the total disease severity at the infected equilibrium reads, according to
\eq{eq:sevtot-def}:
\begin{equation}\label{eqapap:sevtot-ps}
y_\mrm{tot}^*(n) = r \frac{ \left( \beta_\mrm{d} + (n-1) \beta_\mrm{nd}
    \right) K_\mrm{tot} - \mu n } {n  \mu (\mu - r) + r K_\mrm{tot} \left( \beta_\mrm{d} + (n-1) \beta_\mrm{nd} \right)}.
\end{equation}

Next, we take the limit of very large $n$ in
\eq{eqap:sevtot-ps}, which yields
\begin{equation}\label{eq:sevtot-ps-limn}
y_\mrm{tot}^*(n)_{n \to \infty} = r \frac{ R_\mrm{0nd} - 1 } { \mu  + r (R_\mrm{0nd} - 1)},
\end{equation}
where 
\be\label{eq:brn-nd}
R_\mrm{0nd} = b_\mrm{nd} K_\mrm{tot} / \mu
\ee
is the basic reproductive number of pathogen strains as a whole in the
absence of their preferred hosts.

\end{document}